\documentclass[aps,prb,twocolumn,floats]{revtex4-2}
\usepackage{amsmath}
\usepackage{amsfonts}
\usepackage{amssymb}
\usepackage{color}
\usepackage{graphicx} 
\graphicspath{ {/} }
\newcommand{\br}{{\bf r}}
\newcommand{\STO}{SrTiO$_3${ }}
\newcommand{\LAO}{LaAlO$_3${ }}
\newcommand{\LAOSTO}{LaAlO$_3$/SrTiO$_3${ }}

\begin{document}
\title{Influence of a Realistic Multiorbital Band Structure on Conducting Domain Walls in Perovskite Ferroelectrics}
\author{B.C. Cornell}
\email[]{brennancornell@trentu.ca}
\author{W.A. Atkinson}
\email[]{billatkinson@trentu.ca}
\affiliation{Trent University, Department of Physics and Astronomy, Peterborough, Ontario K9L 0G2, Canada}
\date{\today}

\begin{abstract}
Domain wall morphologies in ferroelectrics are believed to be largely shaped by electrostatic forces.  Here, we show that for conducting domain walls, the morphology also depends on the details of the charge-carrier band structure.  For concreteness, we focus on transition-metal perovskites like BaTiO$_3$ and SrTiO$_3$.  These have a triplet of $t_{2g}$ orbitals attached to the Ti atoms that form the conduction bands when electron doped.  We solve a set of coupled equations---Landau-Ginzburg-Devonshire (LGD) equations for the polarization,  tight-binding Schr\"odinger equations for the electron bands, and Gauss' law for the electric potential---to obtain polarization and electron density profiles as a function of electron density.  We find that at low electron densities, the electron gas is pinned to the surfaces of the ferroelectric by a Kittel-like domain structure.  As the electron density increases, the domain wall evolves smoothly through a zigzag head-to-head structure, eventually becoming a flat head-to-head domain wall at high density.  We find that the Kittel-like morphology is protected by orbital asymmetry at low electron densities, while at large electron densities the high density of states of the multiorbital band structure provides effective screening of depolarizing fields and flattens the domain wall relative to single-orbital models. Finally, we show that in the zigzag phase, the electron gas develops tails that extend away from the domain wall, in contrast to na\"{i}ve expectations.
\end{abstract}

\maketitle

\section{Introduction}
\label{Introduction}
Domain wall formation is almost unavoidable in ferroelectric materials because of the strong depolarizing electric fields generated by the spontaneous polarization.  In recent years, focus has shifted away from the macroscopically averaged impact of domains and towards individual domain walls themselves \cite{MeierReview:2022}.  This shift is motivated by successful demonstrations that domain walls may act as reconfigurable nanodevices, for example memristors \cite{mcconville_ferroelectric_2020}, nonvolatile memory \cite{Sharma:2017}, or logic units \cite{Wang:2022}.  Key to these developments was the observation that  domain walls may be made conducting in a number of ferroelectric materials \cite{bednyakov2018physics}.

Domain wall conductivity is the result of the two-dimensional (2D) bound charge density $\sigma_\mathrm{DW}$ that is intrinsic to boundaries separating domains with  different polarizations, ${\bf P}_1$ and ${\bf P}_2$, namely
\begin{equation}
\sigma_\mathrm{DW} = ({\bf P}_1 - {\bf P}_2)\cdot \hat{\bf n}_1,
\end{equation} 
where $\hat {\bf n}_1$ is the outward normal unit vector for domain 1.  Because domain walls typically form 2D sheets, electric fields due to $\sigma_\mathrm{DW}$ tend to be long-range and disruptive to ferroelectricity.  In most cases, neutral domain walls, with $\sigma_\mathrm{DW} = 0$, are energetically preferred.  However, if compensating charges---such as itinerant electrons or holes, or mobile oxygen vacancies---are available, they may collect at charged domain walls and screen long-range fields \cite{bednyakov2018physics}, thus stabilizing the charged domain walls.  When the compensating charges are mobile, the domains form 2D conducting channels that may be manipulated by, for example, external electric fields \cite{McGilly:2015,Oh:2015,Li:2016,Ma:2018,Risch:2022}.

There were a number of early theory papers that proposed mechanisms for the formation of charged domain walls \cite{guro:1968,krapivin:1970,guro:1970,vul:1973}, but the field  only took off much later, following the observation of conduction along head-to-head domain walls in BiFeO$_3$ \cite{Seidel:2009,Seidel:2010,Zhang:2019}.   Since then, charged domain walls have been observed in several proper \cite{SlukaTomas2013Fgac,godau2017enhancing,Crassous:2015,Risch:2022} and improper \cite{choi2010insulating,meier2012anisotropic,Wu:2012,Oh:2015,Wu:2018} ferroelectrics.  From theory considerations, it was argued that without extrinsic influences, charged domain walls in proper ferroelectrics are energetically unstable \cite{gureev2011head,sturman2015quantum}; that is, the energy to produce electron-hole pairs in sufficient numbers to screen $\sigma_\mathrm{DW}$ is larger than the energy gained by forming the domain wall.  Extrinsic stabilizing elements include surfaces that pin the polarization, donor or acceptor impurities, and an external charge reservoir; the latter two of these mechanisms reduce the energetic cost to form a compensating electron or hole gas \cite{nataf2020domain}.  

Two other issues that have been discussed at some length are the width of, and net charge on, conducting domain walls.  Gureev \textit{et al.}\ \cite{gureev2011head} and Sturman \textit{et al.}\ \cite{sturman2015quantum} predicted that the conducting domain wall width is roughly an order of magnitude longer than that of neutral domain walls and depends on the electron or hole effective mass.  Conceptually, this point is important as it shows that the compensating electron or hole gas is an equal partner to the polarization in determining domain wall properties.  The net domain wall charge is also important as it  determines the response of the domain wall to an applied field \cite{gureev2012ferroelectric}.  Na\"{i}vely, one expects a positively charged domain wall to move in the direction of an applied field; however, the situation can be more subtle and Chapman \textit{et al.}\ \cite{chapman2022mechanism} found in their simulations that a flat head-to-head domain wall moves oppositely to the applied electric field, yielding an apparent negative dielectric response.

All of the theoretical calculations reported above assumed that the domain wall has a flat 2D geometry, and a rather different picture emerges when this assumption is relaxed.  In Ref.~\cite{atkinson_domain_structure2022}, it was shown that in a thin ferroelectric film, there is a smooth evolution from lamellar ``Kittel'' domains---that is, alternating domains with opposite polarization, separated by neutral domain walls---at vanishing electron density, to a single flat head-to-head domain wall  at high electron densities.  At intermediate electron densities, one obtains zigzag domain walls.  This evolution, and in particular the zigzag morphology, is driven primarily by imperfect electrostatic screening of the domain wall charge.  Similar considerations led Marton \textit{et al.}\ \cite{marton2023zigzag} to show that randomly distributed charged impurities will also generate a zigzag domain wall.

Here, we move beyond purely electrostatic considerations and explore what happens when the compensating charge is hosted by a realistic multiorbital band structure.  Ref.~\cite{atkinson_domain_structure2022} assumed that $\sigma_\mathrm{DW}$ was compensated by itinerant electrons with an isotropic effective mass.  However, in transition-metal perovskites, which have chemical formula ABO$_3$, the conduction bands are formed from B-cation $d$ orbitals with $t_{2g}$ symmetry.  This immediately raises two questions:  how does the band structure affect the shape of the domain walls, and how does the domain wall shape affect the electron density in the individual orbitals?

To address these questions, we adapt the model used in Ref.~\cite{atkinson_domain_structure2022} to include conduction bands derived from  $t_{2g}$ orbitals.  This model is based on perovskite bilayers in which a 2D electron liquid (2DEL) forms spontaneously at the interface between two otherwise-insulating perovskites. The best-known example of this is the LaAlO$_3$/SrTiO$_3$ bilayer, which becomes conducting when the LaAlO$_3$ cap layer exceeds a few monolayers in thickness \cite{Ohtomo:2004hm,Thiel:2006eo} due to a spontaneous electron transfer from the \LAO surface to the SrTiO$_3$ side of the interface \cite{Bristowe:2014fc}.  These interfaces are interesting for several reasons.  First, \STO can be made ferroelectric by the substitution of Ca \cite{bednorz84} or Ba \cite{lemanov:1996} for Sr, or by the application of strain \cite{Uwe:1976fj,Haeni:2004gj}.  Indeed, several groups have grown ferroelectric \STO interfaces \cite{Bark:2011fo,zhou_artificial_2019,brehin_switchable_2020,Tuvia:2020} and, importantly, Tuvia \textit{et al.}\ \cite{Tuvia:2020} demonstrated hysteretic control of current through their device.  Second, the \LAOSTO system is tunable; both the electron density and its spatial distribution can be modified by gating, while the ferroelectric polarization can be tuned by changing the chemical composition or strain.  Third, considerable effort has been made to control oxygen defect formation during sample growth \cite{liu:2013}, so that most of the 2DEL originates from the external charge reservoir, namely the \LAO surface.  Note, however, that although we have chosen a specific model system, the results described herein should apply broadly to electron-doped perovskite ferroelectrics.

Section \ref{Methods} describes the model and calculational approach in detail. As in Ref.~\cite{atkinson_domain_structure2022}, we calculate the polarization, electron density, and electric potential self-consistently by solving a set of coupled equations:  the LGD equations for the polarization, the Schr\"odinger equation for the electron density, and Gauss' law for the potential.  The novel feature of these calculations is that the band structures are explicitly obtained for three orbitals, $d_{xy}$, $d_{yz}$, and $d_{xz}$, per unit cell via a tight-binding Hamiltonian.  Results of this model are reported in Sec.~\ref{Results} and a comparison to experiments is made in Sec.~\ref{Discussion}.  There, we focus on the effect of orbital anisotropy on the self-consistently calculated domain wall structures, and on the effect of domain wall structure on the orbital selectivity of the resulting band structure.  A summary and conclusions are provided in Sec.~\ref{Conclusion}.

\section{\label{Methods}Model and Calculations}

\begin{figure}[t]
\includegraphics[width=\linewidth]{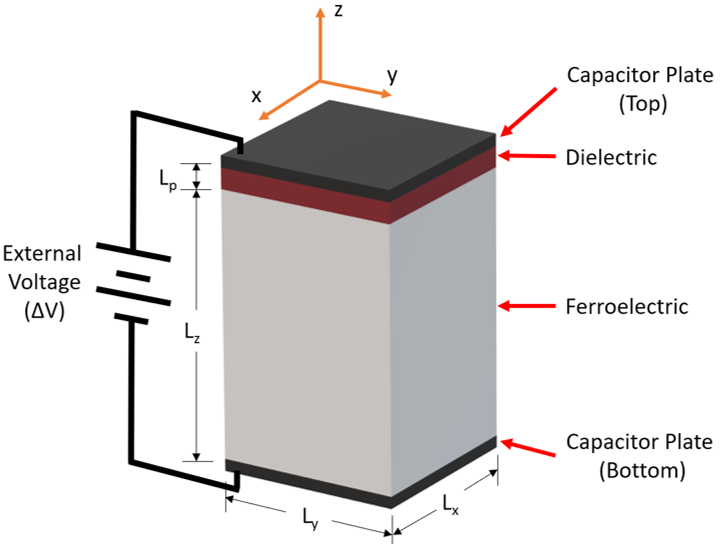}
\caption{\label{Fig:Fig 1 - Model and Polarization} Illustration of the model bilayer.  A ferroelectric substrate (thickness $L_z$) forms a bilayer with a dielectric (thickness $L_p$).  The bilayer is sandwiched between capacitor plates.  The dielectric is insulating, while the ferroelectric is presumed to be electron-doped by a combination of charge transfer from the dielectric and gating by the capacitor plates.  We assume translational invariance along the $y$-axis, and that the domain wall patterns have periodicity $L_x$ in the $x$-direction.}
\end{figure}

As shown in Fig.~\ref{Fig:Fig 1 - Model and Polarization}, the model system comprises a bilayer, with a thin dielectric cap layer (thickness $L_p$) deposited on a thicker ferroelectric substrate (thickness $L_z$).  The entire system is sandwiched between capacitor plates that are maintained at a voltage $\Delta V$.  The system is motivated by \LAOSTO bilayers, in which case the dielectric cap layer represents the \LAO charge reservoir and the substrate represents the \STO film, which is presumed to be made ferroelectric by doping or strain.  The substrate shown in the figure has lateral dimensions $L_x\times L_y$, and we take periodic boundary conditions along the $x$ and $y$ directions.  In experiments, the charge transfer from the reservoir can be modulated by gating; rather than treat this explicitly, our calculations are performed at fixed values of the 2D electron density, $n_\mathrm{2D}$.  For our calculations, then, the cap layer functions as a dielectric that affects the solutions to Gauss' law, but has no effect either on the LGD or Schr\"odinger equations.  It is known that dielectric/ferroelectric bilayers, like that shown in  Fig.~\ref{Fig:Fig 1 - Model and Polarization}, can exhibit an enhanced, so-called negative, capacitance \cite{appleby:2014,khan:2015,lukyanchuk:2018,hoffmann:2021}; this physics is present in the current calculations, but is not directly relevant to our conclusions.

\subsection{Polarization}
The ferroelectric and the polar cap have total polarizations at position $\br$ satisfying
\begin{equation}
\mathbf{P}^\mathrm{total}(\textbf{r})= \mathbf{P}(\textbf{r}) + \mathbf{P}^\mathrm{back}(\textbf{r}), 
\label{eq:total polarization}
\end{equation}
where the ferroelectric polarization, $\mathbf{P}$, arises from the ferroelectric distortion of the unit cell, and the background polarization, $\mathbf{P}^\mathrm{back}$, comes from atomic distortions and non-ferroelectric phonons \cite{levanyuk2016background}.  The background polarization is given by
\begin{align}
\mathbf{P}^\mathrm{back}(\br) &= \epsilon_0\chi(z) \textbf{E}(\br),
\label{eq:background polarization}
\end{align}   
where $\epsilon_0$ is the permittivity of free space, $\textbf{E}$ is the electric field, and $\chi(z)$ is the background dielectric susceptibility.   We have
\begin{align}
\chi(z) = \left\{ \begin{array}{ll}
\chi_\mathrm{FE}, & 0 \leq z \leq L_z \\
\chi_\mathrm{D}, & L_z < z \leq L_z+L_p 
\end{array} \right. .
\end{align}

The ferroelectric polarization $\mathbf{P}$ is obtained by solving a set of LGD equations under the assumption of translational invariance along the $y$-axis (c.f.\ Fig.~\ref{Fig:Fig 1 - Model and Polarization}), so that the polarization is a function of $x$ and $z$ only.  Furthermore, we restrict the polarization to lie in the $x$-$z$ plane, so 
\begin{equation}
\mathbf{P}(\mathbf{r}) = [ \begin{array}{ccc} P_x(x,z),& 0 ,& P_z(x,z) \end{array} ]. 
\end{equation}
We take periodic boundary conditions along the $x$ direction, $\mathbf{P}(x+L_x,z) = \mathbf{P}(x,z)$, and set the derivatives of $\mathbf{P}$ to zero at the top and bottom surfaces of the ferroelectric,
\begin{eqnarray}
\frac{\partial P_x}{\partial z}(x,0) &=& \frac{\partial P_x}{\partial z}(x,L_z) = 0 \label{eq:boundary condition dPx}, \\
\frac{\partial P_z}{\partial z}(x,0) &=& \frac{\partial P_z}{\partial z}(x,L_z) = 0 \label{eq:boundary condition dPz}.
\end{eqnarray}

The ferroelectric polarization is obtained from a fourth-order LGD free energy,
\begin{eqnarray}
{\cal F}_P &=& \int_0^{L_x} dx \int_0^{L_z} dz \Bigg \{ 
\frac {g_{11}}2 \left[ \left ( \frac{\partial P_x}{\partial x}\right ) ^2 + \left ( \frac{\partial P_z}{\partial z}\right ) ^2 \right ] \nonumber \\ 
&& + \frac {g_{44}}2 \left[ \left ( \frac{\partial P_x}{\partial z}\right ) ^2 + \left ( \frac{\partial P_z}{\partial x}\right ) ^2 \right ] 
\nonumber \\
&& + \frac{a_1}{2}P_x^2 + \frac{a_3}{2}P_z^2  + \frac{b}{4} \left( P_x^2 + P_z^2 \right )^2 + \frac{1}{2\epsilon_0}
|\mathbf{P}|^2 
\nonumber \\
&& + \frac{b'}{2} P_x^2P_z^2 \Bigg \}  - \frac{1}{\epsilon_0}\textbf{D}\cdot \mathbf{P}, 
\label{eq:LGD free energy}
\end{eqnarray}
where $\mathbf{D} = \epsilon_0 {\bf E} + \mathbf{P}$ is an electric displacement that contains contributions from the free electrons in the substrate and capacitor plates, and from the background polarization.  Our free energy equation does not contain any terms related to the strain or tilt, which are sometimes included. Most of the parameters used in Eq.~(\ref{eq:LGD free energy}) are for SrTiO$_3$, and are given in Table \ref{table:parameters}.    The parameters $a_1>0$ and $a_3<0$ are chosen so that the spontaneous polarization preferentially aligns with the $z$-axis; this allows us to avoid spurious solutions with the polarization aligned parallel to the surfaces.

Minimizing the free energy with respect to $P_x$ and $P_z$ at fixed $\mathbf{D}$, we obtain  
\begin{align}
\frac{\delta {\cal F}}{\delta P_x} =& P_x \left[a_1 + \frac{1}{\epsilon_0} + b |{\bf P}|^2 + b' P_z^2 \right] 
 - g_{11}\left( \frac{\partial^2 P_x}{\partial x^2} \right) \nonumber \\ 
& - g_{44}\left( \frac{\partial^2 P_x}{\partial z^2} \right) - \frac{1}{\epsilon_0} D_x = 0,
\label{eq:FP_Px} 
\end{align}
and,
\begin{align}
\frac{\delta {\cal F}}{\delta P_z} =& P_z \left[a_3 + \frac{1}{\epsilon_0} + b |{\bf P}|^2 + b' P_x^2  \right]  - g_{44}\left( \frac{\partial^2 P_z}{\partial x^2} \right) \nonumber \\ 
&
- g_{11}\left( \frac{\partial^2 P_z}{\partial z^2} \right) - \frac{1}{\epsilon_0} D_z = 0, 
\label{eq:FP_Pz}  
\end{align}
which are solved for $P_x$ and $P_z$.
In practice, these equations are solved on a discrete grid, with grid spacing $\Delta = 1$~nm.

\begin{table}
\begin{center}
\begin{tabular}{c|c|c}
 Parameter & Value & Units\\
\hline
 $a_1$ & $2 \times 10^8$ & $\mathrm{C}^{-2}\mathrm{m}^2\mathrm{N}$  \\
 $a_3$ & $-1.6 \times 10^8$ & $\mathrm{C}^{-2}\mathrm{m}^2\mathrm{N}$ \\
 $b$ & $5.88 \times 10^9$ & $\mathrm{C}^{-4}\mathrm{m}^4\mathrm{N}$ \\
 $b'$ & $-2.94 \times 10^9$ & $\mathrm{C}^{-4}\mathrm{m}^4\mathrm{N}$ \\
 $g_{11}$ & $2 \times 10^{-10}$ & $\mathrm{C}^{-2}\mathrm{m}^6\mathrm{N}$ \\
 $g_{44}$ & $2 \times 10^{-10}$ & $\mathrm{C}^{-2}\mathrm{m}^2\mathrm{N}$ \\
 $\chi_\mathrm{D}$ & 25 & - \\
 $\chi_\mathrm{FE}$ & 4.5 & - \\
\hline
 $t_0$ & 0 & meV \\
 $t_\parallel$ & 236 & meV \\
 $t_\perp$ & 35 & meV \\
\hline
 $a$ (Lattice Constant) & 0.395 &  nm  \\
 $\Delta$ (Grid Spacing) & 1 & nm \\
 $L_x$ & 28 & nm \\
 $L_y$ & 28 & nm \\
 $L_z$  & 46 & nm \\
 $L_p$  & 5 & nm 
\end{tabular}
\end{center}
\caption{Table of model parameters.  LGD parameters are taken from Appendix A in Ref.~\cite{rabe2007physics} for SrTiO$_3$, with the exception of $a_3$, which is chosen to produce a ferroelectric instability.  Tight-binding parameters are taken from Shubnikov-de Haas measurements \cite{Allen:2013wk}.  The background susceptibilities $\chi_D$ and $\chi_{FE}$ are for \LAO and \STO respectively, while the lattice constant $a$ is for \STO. }
\label{table:parameters}
\end{table}

\subsection{Schr\"odinger Equation}
\label{sec:tbmodel}
The free-electron density is obtained from a three-orbital tight-binding Hamiltonian that includes degenerate $d_{xy}$, $d_{yz}$, and $d_{xz}$ orbitals; in cubic \STO or BaTiO$_3$, these orbitals belong to the Ti atoms and make the dominant contribution to the conduction band.  We include only nearest-neighbour hopping, and keep only the largest hopping matrix elements.  Furthermore, we ignore spin-orbit or polarization-dependent contributions that mix the different orbital symmetries.  Our resulting Hamiltonian $\mathbf{H}$ is therefore block diagonal in the orbital type $\alpha = xy,\, yz,\, xz$.  Translational invariance along the $y$-axis allows us to Fourier transform the Hamiltonian along that dimension, so that we have a mixed representation $(i_x, k, i_z)$, with $(i_x,i_z)$ specifying a spatial location in the $x$-$z$ plane and $k$ representing the wavevector along the $y$-axis.

We write the matrix elements of the tight-binding Hamiltonian $\textbf{H}^\alpha$ as
\begin{align}
H^{\alpha}_{IJ}(k) = \left\{ 
\begin{array}{ll}
t^{\alpha}_0 - e\phi_I + 2t^{\alpha}_y\cos(ka), & i_x=j_x, i_z=j_z \\
t^{\alpha}_x, & i_x = j_x\pm1,i_z = j_z \\
t^\alpha_z, &  i_x=j_x, i_z = j_z\pm 1 \\
0, &\mathrm{ otherwise} 
\end{array} \right. 
\label{eq:Hterms}
\end{align}   
where $\phi_I$ is the electric potential at lattice site $I$, $-e$ is the electron charge, $a$ is the lattice constant, and  $(i_x,i_z)$ and $(j_x,j_z)$ are the $x$- and $z$-coordinates for lattice points I and J respectively.   The parameters and $t^\alpha_0$ and $t^\alpha_w$, $w=\{x,y,z\}$,  are the on-site and nearest-neighbor-hopping matrix elements.  
We assume that the unit cell has cubic symmetry, so that 
\begin{equation}
t^{{xy}}_0 = t^{{xz}}_0 = t^{{yz}}_0 = t_0.
\label{eq:Hterm - t0}
\end{equation}
Cubic symmetry dictates that there are only two distinct nearest-neighbor hopping matrix elements,
\begin{eqnarray}
t^{xy}_x &=& t^{xy}_y = t^{yz}_y = t^{yz}_z = t^{xz}_x = t^{xz}_z = -t_\|, \\
t^{xy}_z &=& t^{yz}_x = t^{xz}_y = -t_\perp.
\end{eqnarray}
Values for the tight-binding parameters are given in Table~\ref{table:parameters}.

The hopping matrix elements in $H^\alpha_{IJ}(k)$ ouple $t_{2g}$ orbitals belonging to neighboring unit cells, separated by a lattice constant $a\approx 4$~\AA.  Diagonalization of the Hamiltonian matrices is, by far, the slowest step in these calculations.  To study physically interesting system sizes, therefore, we coarse-grain the Hamiltonian on a grid, with gridpoints spaced by $\Delta = 1$~nm.  The coarse-graining process preserves the low-energy spectrum and is exact in the limit of low electron densities.  We denote the grid points in the $x$-$z$ plane by $m = (m_x,m_z)$ and $n = (n_x, n_z)$. The coarse-grained Hamiltonian is then
\begin{align}
\tilde H^\alpha_{mn}(k) = \left\{ 
\begin{array}{l}
-e \phi_m -2\tilde t^\alpha_x - 2\tilde t^\alpha_z - 2\tilde t^\alpha_y \\
\hspace{1cm} - 2t^\alpha_y\left(\frac{a^2}{\Delta^2}\right)\cos(k\Delta), \, m=n \\
-t^\alpha_x a^2/\Delta^2,\quad m_x = n_x\pm 1, m_z = n_z \\
-t^\alpha_z a^2/\Delta^2,\quad m_x = n_x, m_z = n_z \pm 1\\
0,\quad \mathrm{ otherwise} 
\end{array} \right. 
\label{eq:CGHamTerms}
\end{align}
where $\tilde t^\alpha_w = t^\alpha_w\left(1 - a^2/\Delta^2 \right)$  and $t_0 = 0$.
We diagonalize this matrix computationally to obtain the eigenenergies $E_{\eta k\alpha}$ and eigenvectors $U_{n \eta}(k\alpha)$, with band index $\eta$, for the free electrons occupying each orbital type $\alpha$ and wavevector $k$.

We can then obtain the electron density at each grid point,
\begin{align}
n_e(n_x,n_z) =& \frac{2}{\Delta^3} \sum_{\eta,k,\alpha} f(E_{mk\alpha})| U_{m \eta}(k\alpha)|^2
\label{eq:freechargedens}
\end{align}   
where the $2$ comes from spin,  $\Delta^3$ is the volume of a single grid point, and $f(\epsilon)$ is the Fermi-Dirac function.  We note that calculations are performed at fixed electron density, which means that the chemical potential $\mu$ must be obtained self-consistently.  We determine $\mu$ by requiring that the 2D electron density is 
\begin{align}
n_\mathrm{2D} = \frac{2}{L_xL_y} \sum_{\eta, k, \alpha} f(E_{\eta k \alpha}).
\label{eq:mu}
\end{align} 
To help stabilize the numerical calculations, we take temperature $T = 10$~K.

For purposes of comparision, we also present results for a one-band \textit{isotropic} model, identical to that used in Ref.~\cite{atkinson_domain_structure2022}.  In the isotropic model, the hopping matrix elements are $t_\| = t_\perp = \hbar^2/2ma^2 = 244$~meV, with $m$ the bare electron mass and $a$ the lattice constant.    

\subsection{Electric Potential}
Given the electron density and polarization, we can calculate the electric potential, $\phi(\br)$ across the lattice using the differential form of Gauss' law,
\begin{equation}
-\nabla^2 \phi = \frac{\rho(\br)}{\epsilon_0},
\label{eq:phi}
\end{equation}
where $\rho(\br)$ is the sum of the free charge density $\rho^f(\br)=-en_e(\br)$, the bound charge density 
\begin{align}
\rho^b(\br) = - \nabla \cdot \mathbf{P}(\br) -\nabla \cdot \mathbf{P}^\mathrm{back}(\br),
\label{eq:bound charge}  
\end{align}
and subject to the boundary conditions
\begin{align}
\phi(x,0) &= 0, \label{eq:boundary condition phi = 0 at bottom}\\
\phi(x,L_z + L_p) &= \Delta V. \label{eq:boundary condition phi = 0 at top}
\end{align} 
The electric field, $\mathbf{E}$, is determined by
\begin{align}
\mathbf{E}(\br) =& -\nabla\phi(\br). 
\label{eq:EfmPhi}
\end{align}

\subsection{Numerical Approach}
In a single iterative loop, we first solve Eqs.~(\ref{eq:FP_Px}) and (\ref{eq:FP_Pz}) to find $P_x$ and $P_z$ for a fixed displacement $\mathbf{D}$.  We then numerically diagonalize Eq.~(\ref{eq:CGHamTerms}) to obtain the eigenvectors and eigenenergies of the coarse-grained tight-binding Hamiltonian for a fixed potential $\phi(m_x,m_z)$.  The chemical potential is then determined from Eq.~(\ref{eq:mu}) for a fixed value of $n_\mathrm{2D}$, and the local electron density is obtained from Eq.~(\ref{eq:freechargedens}).  Finally, we solve  Eqs.~(\ref{eq:phi}) and (\ref{eq:EfmPhi}) for the electric potential and electric field for fixed polarization and electron density.  This cycle is repeated until convergence is obtained.

In general, convergence is significantly more difficult to achieve when free electrons are present in the lattice.  We have found that it is usually better to solve the LGD equations at fixed $\mathbf{D}$ than fixed $\mathbf{E}$; and we have used both Anderson mixing and simple mixing of the polarization and electron density to help with convergence.

\begin{figure*}[tb]
	\centering	
	\includegraphics[width=\linewidth]{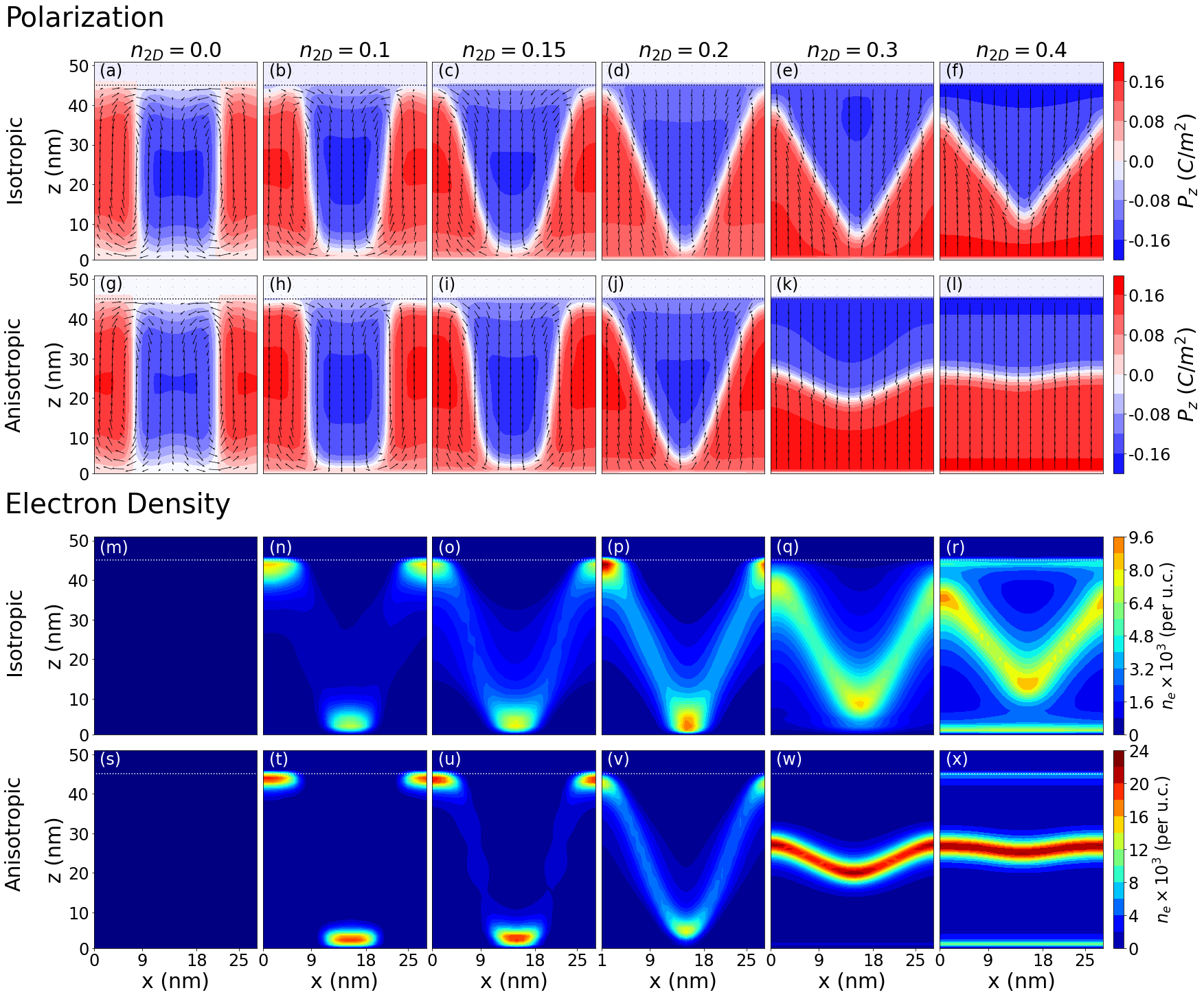}
	\caption{Polarization $\mathbf{P}(x,z)$  and electron density $n_e(x,z)$.  Polarization results are shown for (a)-(f) the isotropic  and (g)-(l) the three-orbital anisotropic  models for six values of the average 2D electron density, $n_\mathrm{2D}$.  Arrows represent the orientation and magnitude of the polarization vector, and the color represents the $z$-component, $P_z$.  The corresponding electron densities for (m)-(r) the isotropic  and (s)-(x) the anisotropic  models are also shown.  The two models are qualitatively similar, but have clear quantitative differences in the domain wall tilts and electron  spatial distributions.}
	\label{fig:Collective Pol and CD}
\end{figure*} 

\section{\label{Results}Results}

\subsection{\label{sec:polarization}Polarization and Electron Density}

We have calculated the polarizations and electron densities as a function of position for a range  of $n_{2D}$ between 0 and 0.4 electrons per 2D unit cell.  For reference, Hall measurements typically report $n_\mathrm{2D}\sim 1\mathrm{-}10 \times 10^{13}$~cm$^{-2}$ ($0.016 \mathrm{-} 0.16$ per 2D unit cell) for \LAOSTO bilayers \cite{pai_physics_2018}, and  $n_\mathrm{2D} \sim 3\times 10^{14}$~cm$^{-2}$ ($0.5$ per 2D unit cell) for GdTiO$_3$/\STO bilayers \cite{moetakef2011electrostatic}.  Furthermore, we choose LGD parameters such that the bulk polarization (neglecting depolarizing fields) is $P_0 = \sqrt{-a_3/b} = 0.165$~C/m$^2$.  This is large relative to the observed polarization in Ca-doped \STO \cite{Rischau:2017vj}, but is consistent with compressively strained \STO \cite{Bark:2011fo}.  We present results for a fixed periodicity, $L_x$, of the domain wall pattern. This is sufficient for us to explore the interplay between multiorbital physics and domain-wall geometry.  However, we expect the optimal domain-wall periodicity to depend on both the thickness of the film \cite{bennett:2020} and $n_\mathrm{2D}$ \cite{atkinson_domain_structure2022}.  

\begin{figure*}[tb] 
	\centering	
	\includegraphics[width=\textwidth]{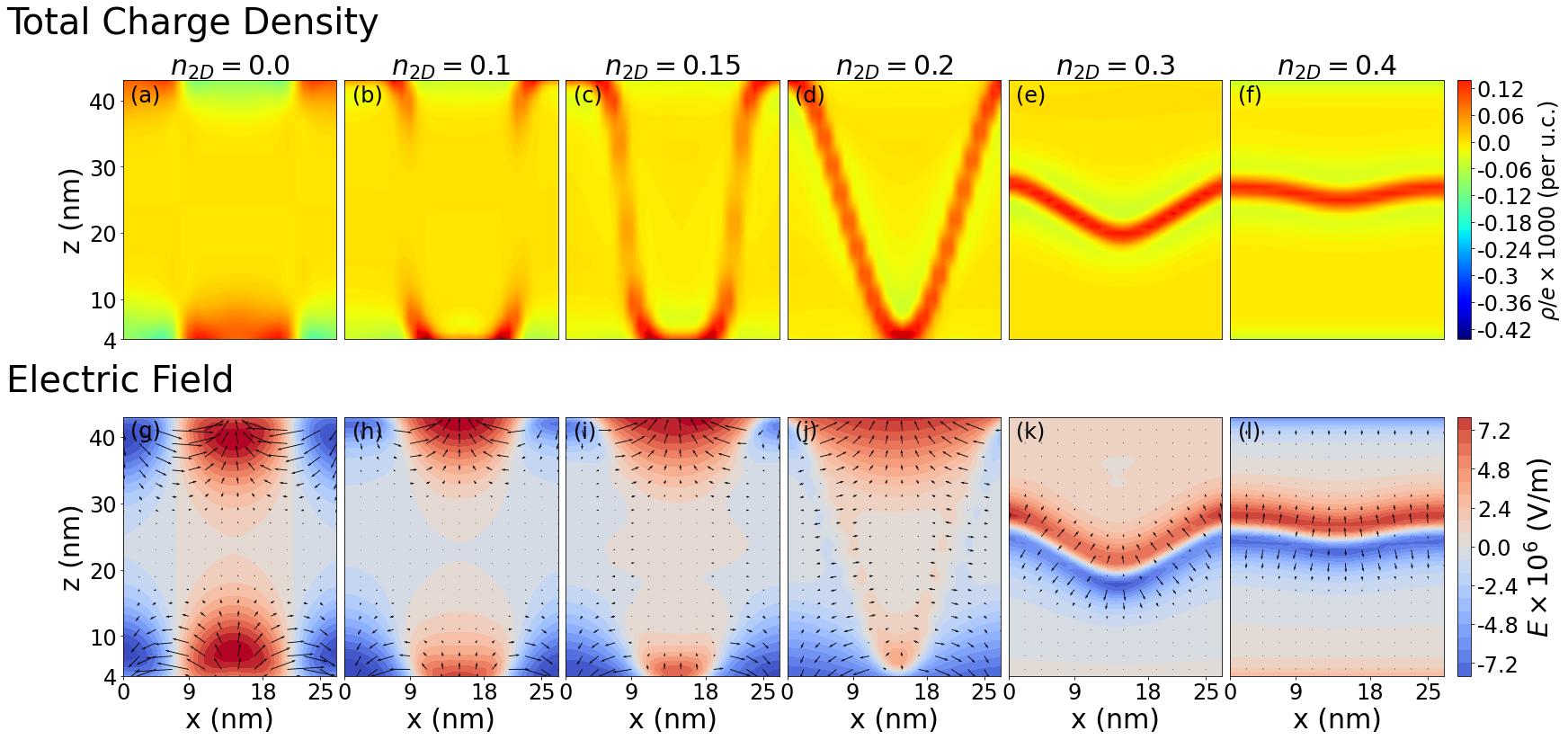}
	\caption{Total charge density and electric field profiles for the anisotropic model.  (a)-(f) The total (bound plus free) charge density is plotted as a function of position for six values of $n_\mathrm{2D}$.  To facilitate comparison with Fig.~\ref{fig:Collective Pol and CD}, we plot $\rho(x,z)/e$.  (g)-(l) The electric field profiles are shown for the same $n_\mathrm{2D}$ values; arrows represent the magnitude and direction of the electric field, and the colormap shows the z-component.  Note that we have removed regions near the top and bottom surfaces of the ferroelectric, where the charge densities and electric fields can be large and overwhelm the plots.}
	\label{fig:ChargeAndField}
\end{figure*}

Figure~\ref{fig:Collective Pol and CD} shows results for both the anisotropic three-orbital model and the isotropic one-band model (See supplemental material for complete results \cite{SI}).  When there are no free electrons  ($n_\mathrm{2D}=0$), the polarization spontaneously breaks up into oppositely polarized domains that are separated by neutral domain walls with $\sigma_\mathrm{DW}=0$ [Fig.~\ref{fig:Collective Pol and CD} (a) and (g)].  These so-called Kittel domains minimize the depolarizing field effects because the bound charge alternates sign along the surfaces [Fig.~\ref{fig:ChargeAndField}(a)], and the electric fields are confined to the surface region [Fig.~\ref{fig:ChargeAndField}(g)].  This leaves the bulk of the ferroelectric isolated from the depolarizing fields. 

As $n_{2D}$ increases, the positive ends of the domains---that is, the ends towards which the polarization points---shrink and move inwards from the top and bottom surfaces of the ferroelectric.  The inward motion is apparent even for $n_{2D}=0.1$, where the polarization clearly points inwards everywhere along the surfaces.  As the positive domain ends shrink, the vertical domain walls tilt to form ``arms'' that connect the domain ends.   The tilts are noticeably smaller  for the anisotropic three-orbital model [Figure~\ref{fig:Collective Pol and CD}(h) and (i)] than for the isotropic single-orbital model [Figure~\ref{fig:Collective Pol and CD}(b) and (c)] when  $n_\mathrm{2D}\leq 0.15$.

By $n_{2D}=0.2$, the positive domain ends have shrunk to become vertices of a single zigzag domain wall. This is true for both the isotropic and anisotropic models [Figure~\ref{fig:Collective Pol and CD}(d) and (j)].  Thereafter, the domain wall vertices move inwards from the ferroelectric surfaces as $n_\mathrm{2D}$ increases.  When $n_\mathrm{2D} > 0.2$, the difference between the isotropic and anisotropic models is stark:  the isotropic model retains a pronounced zigzag structure up to $n_\mathrm{2D} = 0.4$ [Fig.~\ref{fig:Collective Pol and CD}(e) and (f)], while the anisotropic model rapidly approaches a flat head-to-head configuration [Fig.~\ref{fig:Collective Pol and CD}(k) and (l)].  We conclude that the multiorbital band structure has a strong quantitative effect on the domain wall morphology.

For the most part, the conduction electrons are bound to the positive domain ends at low electron densities ($n_\mathrm{2D} \leq 0.2$), and progressively spread out along the domain wall as it becomes flatter  [Fig.~\ref{fig:Collective Pol and CD} (m)-(x)]. At high electron densities ($n_\mathrm{2D} = 0.4$) the domain walls are saturated with electrons, and the excess  spills over to the surfaces [Fig.~\ref{fig:Collective Pol and CD} (r), (x)].  
This doping-dependence is true for both the isotropic and anisotropic models, and is expected from electrostatic considerations alone \cite{atkinson_domain_structure2022}.   It is commonly assumed that conducting domain walls are overall charge neutral, which requires that the compensating electron density match the bound charge density.  Figure~\ref{fig:ChargeAndField} shows that this is largely true:  the total charge density (free plus bound) is generally two orders of magnitude smaller than the free or bound charge density alone.  The residual domain wall charge is always positive in our calculations.  Interestingly, Fig.~\ref{fig:ChargeAndField} shows that as a result of the residual charge distribution, the electric field  is largely confined to the surfaces for $n_\mathrm{2D}\leq 0.2$ and to the domain walls for $n_\mathrm{2D} > 0.2$.  

Although itinerant electrons screen the bound charge effectively in both models, Fig.~\ref{fig:Collective Pol and CD}  shows that the electrons are much more tightly bound to the  surfaces and domain walls for the anisotropic model.  This can be attributed to two differences between the models.  First, the isotropic conduction band is derived from a single orbital per unit cell, while the anisotropic model has three $t_{2g}$ orbitals per unit cell, each of which can accommodate electrons. As a result, the Fermi energy  is lower in the anistropic model for a given $n_\mathrm{2D}$. Electrons in the anistropic model therefore have a lower kinetic energy and are more tightly bound.

Second, in the isotropic model, the conduction band effective mass is isotropic and equal to the bare electron mass $m$; in the anisotropic model,  each $t_{2g}$ band has a heavy axis with effective mass $m^\ast_\perp \sim 10 m$, and two light axes with $m^\ast_\| \approx m$.   In the bulk, cubic symmetry is preserved for the anisotropic model because the $t_{2g}$ bands are related by point group operations of the cubic lattice.  However surfaces and domain walls both break the cubic symmetry, and therefore the $t_{2g}$ orbital degeneracy.  This leads to a so-called orbital selectivity, in which one band may be preferentially occupied over the others.  In the next section, we discuss how these features of the multiorbital band structure affect the domain wall evolution.

\subsection{\label{sec:orbital selectivity}Influence of the Multiorbital Band Structure}

\begin{figure}[tb]   
	\includegraphics[width=\columnwidth]{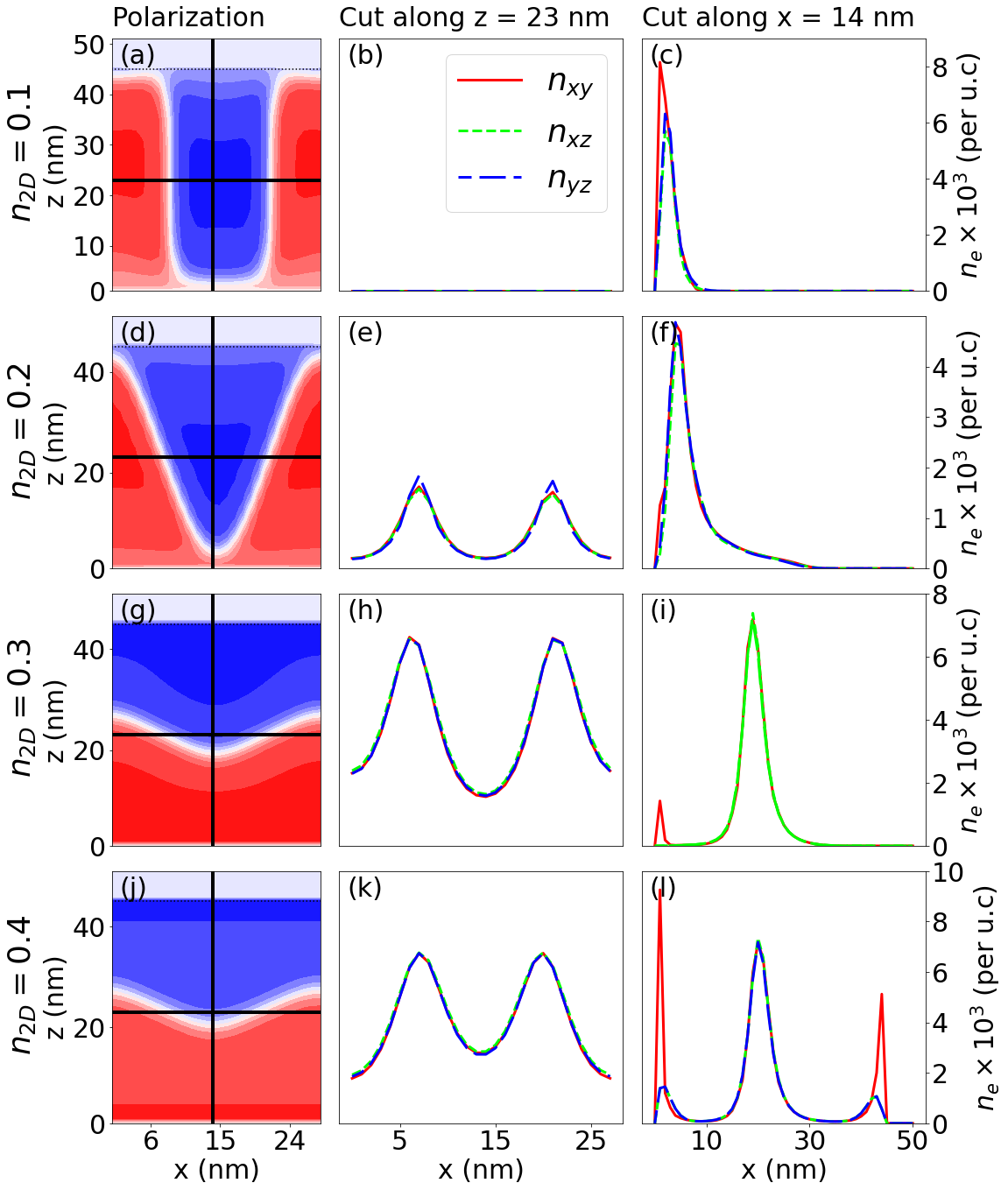}
	\caption{Electron densities plotted for different orbitals types along fixed x and z values.  The first column is a plot of the $z$-component of the polarization [as in Fig ~\ref{fig:Collective Pol and CD} (h)--(l)].  The black lines show the cuts along which the electron densities are plotted.  Columns 2 and 3 show the electron densities for each orbital type along  $z = 23$~nm and $x = 14$~nm, respectively.}  
	\label{fig:Collective Charge Density Differences}
\end{figure}

Orbital selectivity has been discussed at length in the context of non-ferroelectric \LAOSTO bilayers \cite{Khalsa:2012fu,herranz2015engineering,raslan2017temperature}.  In these systems, the interface breaks the  cubic symmetry of the \STO substrate; electrons  with $d_{xy}$ character are heavy along the $z$-axis, perpendicular to the interface, and are therefore easily pinned to the interface by even weak confining potentials.  The lowest energy bands therefore have $d_{xy}$ character and extend only a few unit cells into the substrate.  Conversely, electrons with $d_{xz}$ or $d_{yz}$ character are an order of magnitude lighter along the $z$-axis and extend much farther into the bulk.   

\begin{figure}[tb] 
	\includegraphics[width=\linewidth]{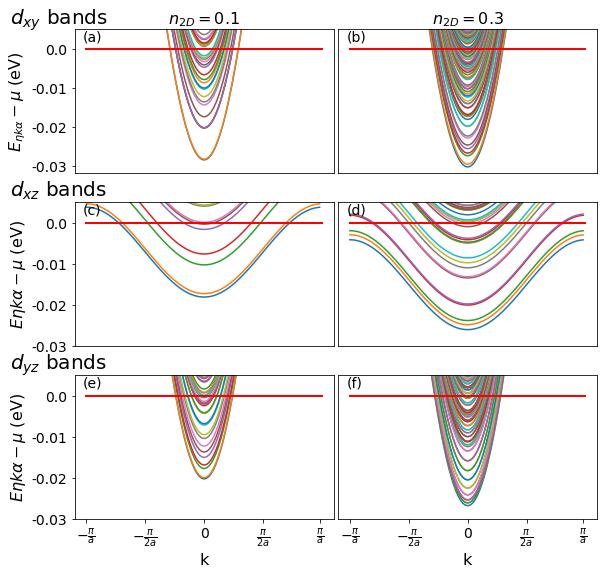}
	\caption{Band structure of the anisotropic model.  Results are shown for (a), (c), (e) $n_\mathrm{2D}=0.1$ and (b), (d), (f) $n_\mathrm{2D}=0.3$.  The red dashed horizontal line indicates the chemical potential.}
	\label{fig:en for n2D=0.1}
\end{figure}

A similar, although more nuanced, situation rises for the ferroelectric case.  Figure~\ref{fig:Collective Charge Density Differences} shows orbitally resolved electron densities along a pair of cuts through the ferroelectric films for a range of $n_\mathrm{2D}$.  When $n_\mathrm{2D} = 0.1$ [Fig.~\ref{fig:Collective Charge Density Differences}(a)-(c)], we see that electrons are confined to the positive ends of the Kittel domains, where they partially compensate the bound charge [Fig.~\ref{fig:Collective Charge Density Differences}(c)].     As in the nonferroelectric case, there is a noticeable orbital selectivity, with the $d_{xy}$ bands more highly occupied and more tightly bound to the interface.   This  is  reflected in the band structure (Fig.~\ref{fig:en for n2D=0.1}), which shows that at $n_\mathrm{2D} = 0.1$ the lowest-energy $d_{xy}$ bands (there is a degenerate pair) lie $\sim 10$~meV below a dense spectrum of $d_{xy}$, $d_{yz}$, and $d_{xz}$ bands.  These two lowest-energy bands correspond to the surface states responsible for the orbital selectivity in Fig.~\ref{fig:Collective Charge Density Differences}(c).

We propose that these surface states, which are unique to the anisotropic model, help stabilize the Kittel-like domain-wall structure against tilting for $n_\mathrm{2D} < 0.2$.  The key idea is that the positive domain ends necessarily shrink as the domain walls tilt away from vertical (c.f.\ Fig.~\ref{fig:Collective Pol and CD}).  As the domain ends shrink, the surface electrons are confined to a smaller volume, which raises their kinetic energy.  There is thus an energetic cost for the domain walls to tilt. This becomes less relevant as $n_\mathrm{2D}$ increases because the potential confining the electrons to the surfaces becomes increasingly screened, which lowers the cost to tilt the domain walls.  While it is difficult to establish cause-and-effect, we propose that the $d_{xy}$ surface states are the key difference between the isotropic and anisotropic models that differentiates the rate of domain-wall tilt at low $n_\mathrm{2D}$.  At sufficiently high electron densities, $en_\mathrm{2D} =0.20 \approx P_0$, the screening is sufficient that the surface electrons escape and the domain wall takes on a zigzag structure that is similar to that of the isotropic model.

Indeed, the $d_{xy}$ surface states are nearly gone by $n_\mathrm{2D} = 0.2$ [Fig.~\ref{fig:Collective Charge Density Differences}(d)-(f)].  There is a small excess of $d_{yz}$ electrons along the diagonal arms of the domain walls; however, orbital selectivity is largely irrelevant once $n_\mathrm{2D} \geq 0.2$ [Fig.~\ref{fig:Collective Charge Density Differences}(d)-(l)].  This is evident in the band structure for $n_\mathrm{2D} =0.3$, where  the lowest-energy bands for each orbital type are nearly degenerate [Fig.~\ref{fig:en for n2D=0.1}(b), (d), and (f)] .  We attribute the weak orbital selectivity along the domain walls to the  large domain wall width, of order 10~nm, and shallow confining potential, of order 30~meV at $n_\mathrm{2D}=0.2$ and smaller at  higher $n_\mathrm{2D}$.  In contrast, electrons that spill over to the surfaces at large
$n_\mathrm{2D}$ have nearly complete $d_{xy}$ character [Fig.~\ref{fig:Collective Charge Density Differences}(i), (l)].  

As remarked above, the largest differences in the domain wall morphologies between the isotropic and anisotropic models occur for $n_\mathrm{2D} >0.2$ [c.f.~Fig.~\ref{fig:Collective Pol and CD}].  Here, the domain walls retain their zigzag structure for the isotropic model but quickly become flat for the anisotropic one.  We believe the difference is simply that the anisotropic model, with three orbitals per unit cell, has a much larger density of states and screens depolarizing electric fields more effectively than the isotropic model.  In the limit of perfect screening, the domain wall should be perfectly flat.

\subsection{\label{sec:domain wall neutrality} Domain Wall Neutrality}

We return to the issue of domain-wall charge neutrality.  As mentioned above, Fig.~\ref{fig:ChargeAndField} shows that the cores of the domain walls have a net positive residual charge, which is at most a few percent of the bound or free charge density.  Closer inspection, however, reveals that the residual charge density does not fall to zero away from the domain walls, but that there is a weak background charge density spread throughout the ferroelectric substrate.  This is especially obvious in Fig.~\ref{fig:Collective Charge Density Differences}(f), where the electron density has a long tail that extends away from the domain wall.

To make sense of this, it is useful to compare our 2D solutions for ${\bf P}(x,z)$ with the simple 1D structure of a flat head-to-head domain wall, for which the polarization is approximately
\begin{equation}
P_z(z) = -P^\mathrm{surf} \tanh\left(\frac{z-z_0}{d}\right)
\end{equation}
where $z_0$ is the domain wall location, $d$ is a length scale that depends on LGD parameters and the electron effective mass \cite{sturman2015quantum}, and where $P^\mathrm{surf}$ is the asymptotic value of the polarization at the surface of the ferroelectric film.  The 2D bound charge density on the domain wall is $\sigma_\mathrm{DW} = 2P^\mathrm{surf}$.  To a good approximation,  
\begin{equation}
P^\mathrm{surf} =\left \{ \begin{array}{ll} 
\frac 12 en_\mathrm{2D}, & en_\mathrm{2D} < 2P_0 \\
P_0, & en_\mathrm{2D} > 2P_0 
\end{array}\right .,
\label{eq:Psurf0}
\end{equation}
where 
$P_0$ is the saturated bulk polarization from the LGD equation in the absence of depolarizing fields, namely,
\begin{equation}
P_0 = \sqrt{\frac{-a_3}{b}}.
\label{eq:P0}
\end{equation}
The first of the two expressions in Eq.~(\ref{eq:Psurf0}) is obtained by insisting that the domain wall accommodate all of the itinerant electrons and be charge-neutral, so that $\sigma_\mathrm{DW} = en_\mathrm{2D} $.  The second expression in Eq.~(\ref{eq:Psurf0}) applies when the electron density exceeds what can be accommodated on the domain wall.  In this case, the polarization saturates at $P_0$, the wall accommodates a 2D electron density $n_\mathrm{DW} = 2P_0/e < n_\mathrm{2D}$, and the excess electrons spill over to the surfaces of the ferroelectric.  (Equation~(\ref{eq:Psurf0}) ignores the fact that the surface electrons will generally modify the surface polarization.)  For both cases in Eq.~(\ref{eq:Psurf0}), the interior of the ferroelectric film is overall charge-neutral.

A similar analysis of $P^\mathrm{surf}$ can be made for the 2D case at $\Delta V = 0$.  We start by writing the expression for the total charge in the ferroelectric substrate as
\begin{align}
Q_\mathrm{total} &= -e\int d^3r\ n(\br)  - \int d^3r\ \nabla\cdot {\bf P}(\br) \nonumber \\
&= -e n_\mathrm{DW} L_xL_y - \oint dA\  \mathbf{P}\cdot \mathbf{\hat n},
\label{eq:Qtot}
\end{align}
where $\mathbf{\hat n}$ is the outward normal vector from the substrate. The integrals are over the interior volume of the ferroelectric, and therefore do not include the surface charges (free and bound).  The 2D electron density $n_\mathrm{DW}$ includes both the itinerant electrons bound to the domain walls and the dilute electron gas that extends away from the domain walls.

Keeping in mind that there are both top and bottom surfaces and that, except at $n_\mathrm{2D} = 0$, the polarization points inwards everywhere along these surfaces, we can generalize the previous definition to obtain the average (inwards) surface polarization,
\begin{equation}
P^\mathrm{surf} = - \frac{1}{2L_xL_y} \oint dA\  \mathbf{P}\cdot \mathbf{\hat n}.
\label{eq:2Psurf}
\end{equation}
Then, as before, overall neutrality of the substrate (i.e.\ $Q_\mathrm{total}=0$) requires $P^\mathrm{surf} = \frac 12 e n_\mathrm{DW}$.  Note, however, that unlike for the flat domain wall, there is no expectation that $P^\mathrm{surf}$ is also the maximum value of the polarization.

\begin{figure}[h] 
	\centering	
	\includegraphics[width=\linewidth]{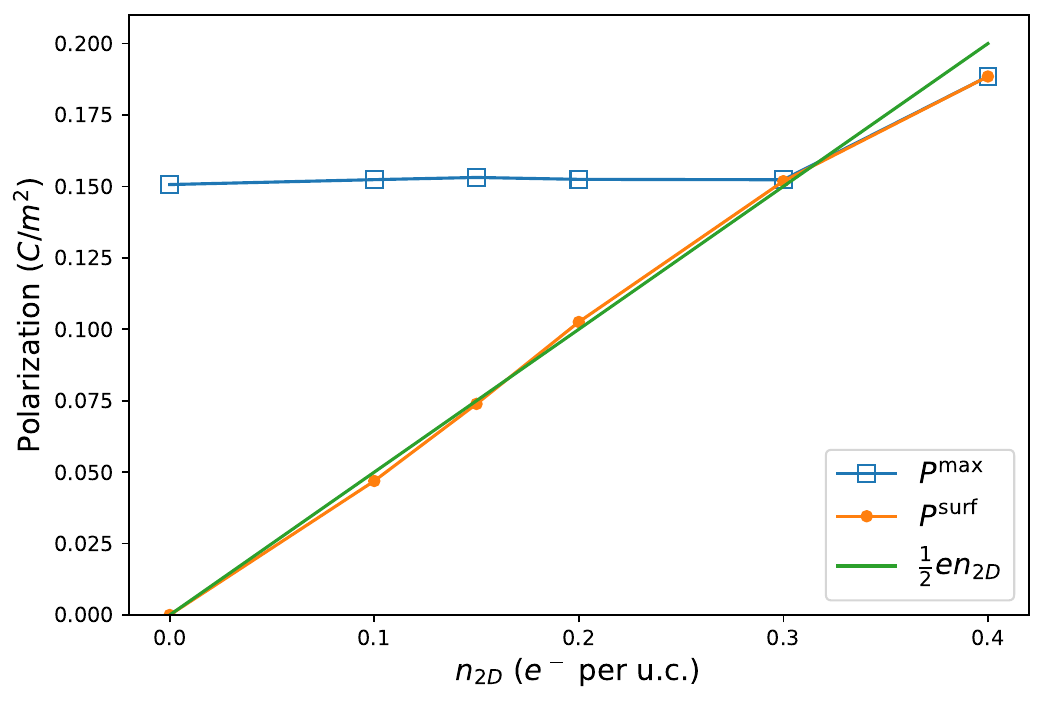}
	\caption{The maximum polarization $P^\mathrm{max}$, the average surface polarization $P^\mathrm{surf}$, and the straight line $P = \frac{1}{2}en_{2D}$ are plotted versus $n_{2D}$ for the anisotropic model.  } 
	\label{fig:SurfaceVsBulkPol}
\end{figure}

Figure~\ref{fig:SurfaceVsBulkPol} shows a plot of $P^\mathrm{surf}$ from Eq.~(\ref{eq:2Psurf}) versus $n_\mathrm{2D}$. For comparison, we also  plot $\frac 12 en_\mathrm{2D}$, which is the predicted value of $P^\mathrm{surf}$ when all electrons are accommodated by the domain wall.  These two agree very well when $n_\mathrm{2D} \leq 0.3$, but deviate when electrons spill over to the surfaces at $n_\mathrm{2D} = 0.4$.  In this regard, the physics of the domain wall patterns shown in Fig.~\ref{fig:Collective Pol and CD}, are consistent with Eq.~(\ref{eq:Psurf0}), which was obtained for the simple 1D case.

Figure~\ref{fig:SurfaceVsBulkPol} also shows $P^\mathrm{max}$, the maximum magnitude of the polarization within the substrate.  Unlike in the 1D case, this is independent of $n_\mathrm{2D}$ at low electron densities, and is essentially equal to the value $P_0 = 0.165$~C/m$^2$ that one  obtains  from Eq.~(\ref{eq:P0}) for the model parameters in Table~\ref{table:parameters}.  Furthermore, one can see from Fig.~\ref{fig:Collective Pol and CD} that, for $n_\mathrm{2D} \leq 0.2$, the maximum polarization occurs near the center of the ferroelectric, rather than at the surfaces.  

It is only when the domain walls are nearly flat, i.e.\ $n_\mathrm{2D} > 0.2$, that we regain the behavior of the 1D case:  that is, the polarization is a maximum at the surfaces, so $P^\mathrm{max} = P^\mathrm{surf}$.  Interestingly, when $n_\mathrm{2D} = 0.4$ our calculations predict that the polarization is enhanced at the surfaces, namely $P^\mathrm{surf} > P_0$.  Similar results were found previously in 1D models \cite{chapman2022mechanism}, and can be attributed to the electrons that spill over to the surfaces at high electron densities.  These generate electric fields that must be screened by polarization gradients. This is an intriguing mechanism for generating surface-enhanced polarization; however, we caution that there are also short-range electron-lattice interactions that are neglected in our LGD model that will tend to suppress the polarization.

In summary, we have identified two distinct regimes in this section.  The first corresponds to electron densities $n_\mathrm{2D} \lesssim 2P_0/e$, and is likely relevant for most electron-doped ferroelectrics.  In this regime 1D (flat) and 2D (zigzag) domain walls are quite different.  In particular, in the 2D case the maximum polarization does not depend on the requirement for charge neutrality in the ferroelectric, but is simply equal to $P_0$.  However, like the 1D case, the average surface polarization is limited by overall charge neutrality requirements, and to a good approximation satisfies $P^\mathrm{surf} = \frac 12 en_\mathrm{2D}$.   

The second regime corresponds to electron densities  $n_\mathrm{2D} \gtrsim  2P_0/e$, and may be relevant to oxide interfaces with weakly ferroelectric substrates---for example Ca-doped or Ba-doped \STO with $P_0 \sim 0.03$~C/m$^2$---and $e n_\mathrm{2D} \sim 0.1$~C/m$^2$.  In this case, domain walls approach the ideal flat head-to-head structure and can be largely understood within approximate 1D models.

\begin{figure*}[tb] 
	\centering	
	\includegraphics[width=\linewidth]{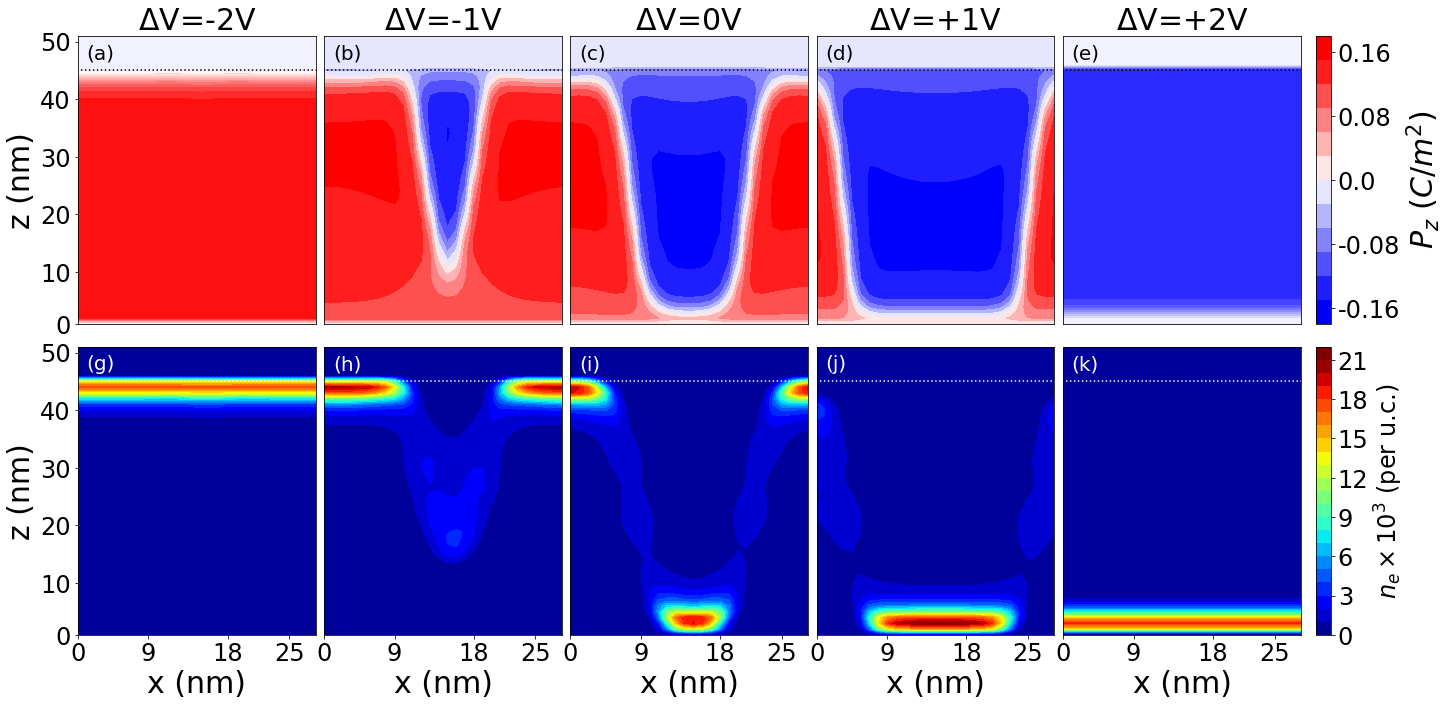}
	\caption{Voltage-dependence of the polarization and electron density for $n_{2D}=0.15$.  Results are for (a)-(e) the $z$-component of the polarization  and (g)-(k) the electron densities. }  
	\label{fig:Bias Voltage for n2D=0.15}
\end{figure*}

\subsection{Dependence on Bias Voltage}

In Sec.~\ref{sec:polarization} and Sec.~\ref{sec:domain wall neutrality}, we showed that the domain walls are approximately neutral, but have a positive residual charge density that is  two orders of magnitude smaller than the bound charge density $\sigma_\mathrm{DW}$ alone.  While small, this residual charge is important as it allows the wall to be manipulated by an external field.  This field is provided by a bias voltage, $\Delta V$, applied between the capacitor plates shown in Fig.~\ref{Fig:Fig 1 - Model and Polarization}, with $\Delta V > 0$ implying that the top plate is at a higher potential.

Figure~\ref{fig:Bias Voltage for n2D=0.15} shows the polarization and total free charge as a function of bias voltage for the case $n_\mathrm{2D}=0.15$.    We observe that the domain wall moves towards regions of lower potential, as might be expected given its net positive charge. This is consistent with the voltage dependence found for zigzag domain walls in the isotropic model \cite{atkinson_domain_structure2022}, but is the opposite of what was obtained previously for head-to-head domain walls with an ideal flat geometry (i.e.\ with translational invariance in the $x$ and $y$ directions), which were found to move against the applied voltage \cite{chapman2022mechanism}.  The discrepancy between two predictions remains to be explained.  

\section{Discussion}
\label{Discussion}

Here, we discuss the relevance of our calculations to recent experiments.  The system described in this work is modeled on the well-known \LAOSTO bilayers, with the \STO made ferroelectric. However, to make a direct comparison to existing experiments, we must note that there is an intrinsic voltage drop across the \LAO layer that is not treated explicitly in our calculations.  This voltage drop results from both the polarity of the \LAO unit cell, as it is typically grown, and of the electron transfer to the \STO interface.  To make a precise statement, one needs to know details of the donor states on the \LAO surface.  However, it is reasonable to guess that the intrinsic bias voltage, $\Delta V_0$, is a substantial fraction of the \LAO band gap, i.e.\ of order $\Delta V_0 \sim -2$~V.   

As we see from Fig.~\ref{fig:Bias Voltage for n2D=0.15}, a negative bias voltage draws electrons towards the top surface and tends to polarize the substrate \textit{upwards}.  There remains, however, a thin layer immediately adjacent to the interface with nearly vanishing polarization. In our model, therefore, the native state of \LAOSTO interfaces is one in which the interface is not representative of the bulk.  We emphasize that this picture applies when $en_\mathrm{2D}< 2P_0$; for larger electron densities, some electrons will spill over from the domain wall to the interface and create a permanent conducting layer there.

The results in Fig.~\ref{fig:Bias Voltage for n2D=0.15} are likely relevant to cases where ferroelectricity is induced by strain.  In Ref.~\cite{Bark:2011fo}, compressively strained interfaces had electron densities $en_\mathrm{2D} \lesssim 0.16$~C/m$^2$, while density functional theory (DFT)  predicted a substrate polarization $P_0 = 0.18$~C/m$^2$, which places this system in the range $en_\mathrm{2D}< 2P_0$.  While no direct measurements of the polarization were made, density functional calculations found that it points away from the interface, into the substrate. This is different from what we find.  However, the calculations in Ref.~\cite{Bark:2011fo} were limited to 5 unit cells of \STO and are therefore unable to capture domain wall structures like those shown in Fig.~\ref{fig:Bias Voltage for n2D=0.15}.  Our results suggest that the compressively strained \LAOSTO system deserves further experimental study.

The experiments of Tuvia \textit{et al.}\ \cite{Tuvia:2020} are also interesting because they showed strong hysteresis in the sheet resistance in their device as a function of gate voltage.  In this case, the \STO substrate was made ferroelectric by Ca-doping, which produced a polarization of order 0.03~C/m$^2$, and relatively low carrier densities with $en_\mathrm{2D} \sim 0.02$~C/m$^2$.  These experiments are therefore also in the regime $e n_\mathrm{2D} < 2P_0$, and should therefore show similar physics as in Fig.~\ref{fig:Bias Voltage for n2D=0.15}.  A complication is that the polarization axis for Ca-doped \STO lies along the $\langle 110 \rangle$ cubic axes, so that head-to-head domain walls will form at 45$^\circ$ angles relative to the cubic crystalline axes.  Indeed, Tuvia \textit{et al.} observed evidence of 1D conducting channels along the $\langle 110\rangle$ directions, which they attributed to structural domain walls associated with octahedral tilts.  We suggest that these might, in fact, be charged domain walls.

Although the model used in this work is motivated by oxide interfaces, the physics is general.  Indeed, zigzag head-to-head or tail-to-tail domain walls have been experimentally observed in established ferroelectrics, such as strained BaTiO$_3$ \cite{Denneulin_2022}.  As found here and in previous simulations \cite{atkinson_domain_structure2022,marton2023zigzag}, zigzag charged domain walls are expected to arise naturally in the presence of compensating charges.  Based on the results shown here, we expect that zigzag domain walls will have a lower itinerant carrier density than the straight walls, and that the itinerant carrier density should be largest at the vertices of the zigzag domain walls.

\section{\label{Conclusion}Conclusion}

In this work, we explored the domain wall structure for an electron-doped ferroelectric film.  Our simulations were for thin (46~nm-thick) films, so that both surface and bulk effects played a role on the domain structure.  We focused specifically on transition-metal perokvskites, such as BaTiO$_3$ or SrTiO$_3$, for which the conduction bands are formed from a triplet of $t_{2g}$ orbitals.  By solving coupled LGD and Schr\"odinger equations, we were able to explore the role of the multiorbital structure in determining both the domain-wall shape and the conduction-band structure.

We found that the general trend with increasing electron density $n_\mathrm{2D}$ is essentially the same as reported in Ref.~\cite{atkinson_domain_structure2022}.  When $n_\mathrm{2D}=0$, the ferroelectric breaks up into Kittel domains, separated by neutral 180$^\circ$ domain walls.  At low electron densities, the Kittel domain structure is preserved, with electrons migrating to the positive ends of the domains at the top and bottom surfaces of the film.  With increasing $n_\mathrm{2D}$, the domain structure evolves continuously through a zigzag head-to-head structure, and finally to a flat charged domain wall.

The main novel feature of the current work is that the zigzag domain wall is less stable in the multiorbital case than in a single-orbital case.  At low $n_\mathrm{2D}$, we find that charged domain wall arms form at higher electron densities than in the single-orbital case.  We attribute this to an orbital selectivity that tightly confines electrons with $d_{xy}$ symmetry to the surfaces, and prevents them from migrating to the connecting arms.  Since charged domain walls require a compensating electron gas, the formation of arms via tilting is suppressed at low $n_\mathrm{2D}$.   At high $n_\mathrm{2D}$, the domain wall evolves much more quickly towards a flat geometry in the multiorbital case than in the single-orbital case.  We attribute this to the high density of states in the multiorbital case, which enhances screening of the depolarizing fields.  Notably, there is no evidence of significant orbital selectivity along the domain walls.  Overall, the zigzag morphology occupies a smaller range of $n_\mathrm{2D}$ values in the multiorbital case.

\begin{acknowledgments}
This work is supported by the Natural Sciences and Engineering Research Council (NSERC) of Canada, and the high performance computing facilities of the Shared Hierarchical Academic Research Computing Network (SHARCNET:www.sharcnet.ca) and Compute/Calcul Canada.
\end{acknowledgments}

\bibliography{Citation,FE_Interface}

\begin{widetext} 
\newpage
\section{\label{supp info}Supplemental Information}

Figures \ref{fig:Results n2D=0.0}-\ref{fig:Results n2D=0.4} show our self-consistent solutions for the polarization ${\bf P}$, orbitally-resolved electron densities ($n_{xy}$, $n_{xz}$, $n_{yz}$), and bound charge density $\rho_b = -\nabla\cdot {\bf P}$.  The color scale in (a) indicates the $z$-component  of the polarization.  

\begin{figure}[b]
	\includegraphics[width=\linewidth]{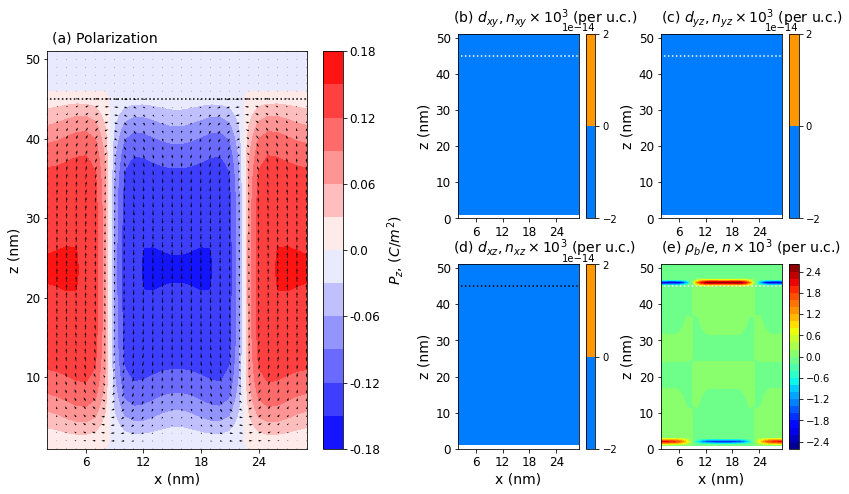}
	\caption{Self-consistent solutions for $n_\mathrm{2D} = 0.0$.}
	\label{fig:Results n2D=0.0}
\end{figure}

\begin{figure}[tb]	
	\includegraphics[width=\linewidth]{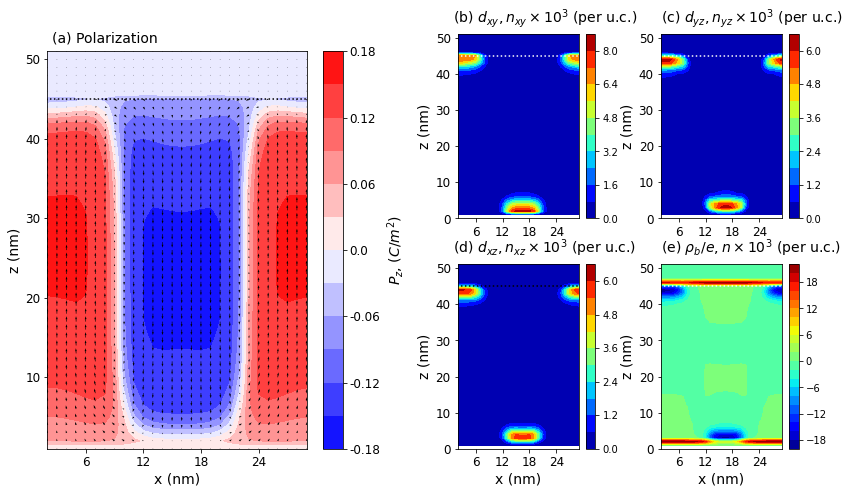}
	\caption{Self-consistent solutions for $n_\mathrm{2D} = 0.10$.}
	\label{fig:Results n2D=0.1}
\end{figure}

\begin{figure}[tb]
	\includegraphics[width=\linewidth]{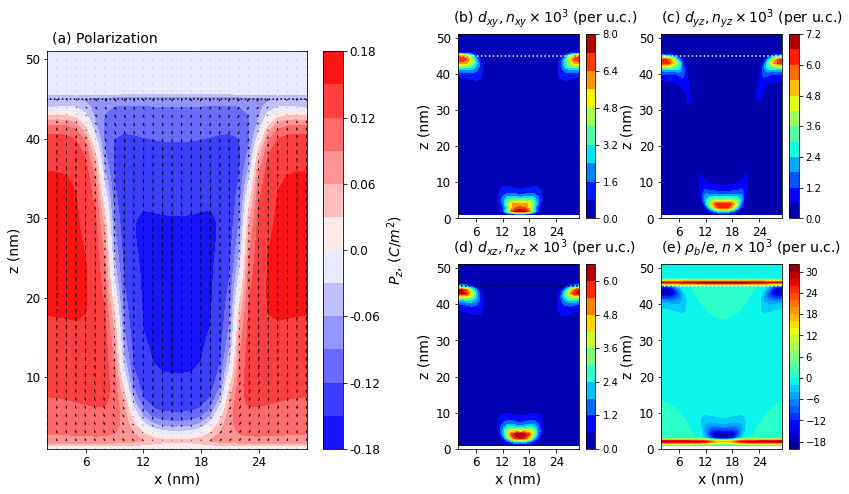}
	\caption{Self-consistent solutions for $n_\mathrm{2D} = 0.15$.  }
	\label{fig:Results n2D=0.15}
\end{figure}

\begin{figure}[tb]
	\centering	
	\includegraphics[width=\linewidth]{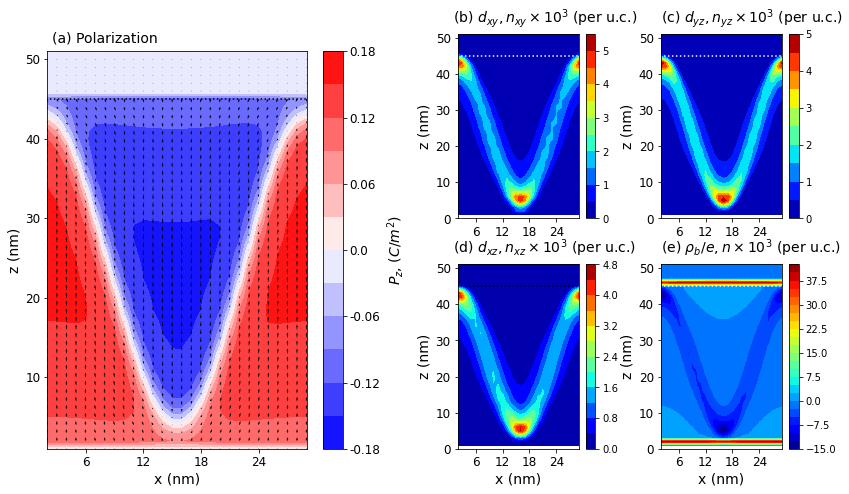}
	\caption{Self-consistent solutions for $n_\mathrm{2D} = 0.20$.}
	\label{fig:Results n2D=0.2}
\end{figure}

\begin{figure}[tb]
	\centering	
	\includegraphics[width=\linewidth]{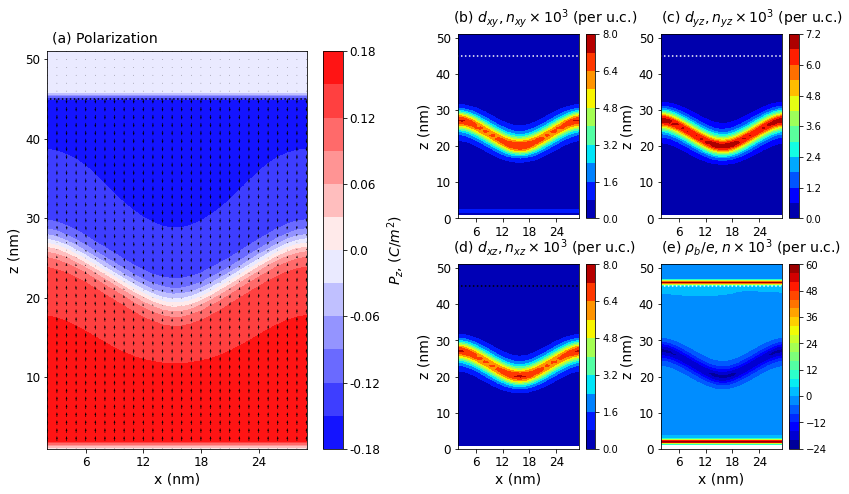}
	\caption{Self-consistent solutions for $n_\mathrm{2D} = 0.30$. }
	\label{fig:Results n2D=0.3}
\end{figure}

\begin{figure}[tb]
	\centering	
	\includegraphics[width=\linewidth]{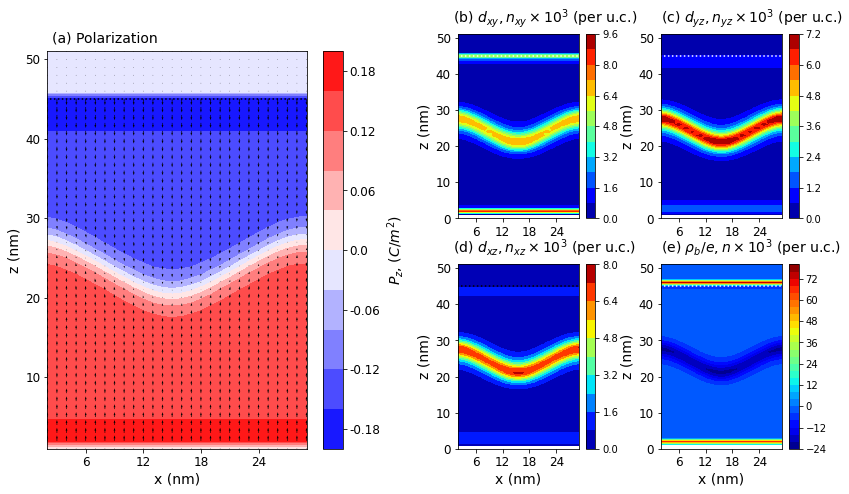}
	\caption{Self-consistent solutions for $n_\mathrm{2D} = 0.40$.}
	\label{fig:Results n2D=0.4}
\end{figure}

\end{widetext}
\end{document}